\documentclass[aps,twocolumn,groupedaddress,showpacs]{revtex4}
\usepackage{longtable}
\usepackage{tabularx}
\usepackage{graphicx}
\usepackage{dcolumn}
\usepackage{bm}
\def\v_op{ \hat{\mathbf v} }

\newcommand{\la}{\langle}
\newcommand{\ra}{\rangle}

\newcommand{\be}{\begin{equation}}
\newcommand{\ee}{\end{equation}}
\newcommand{\bea}{\begin{eqnarray}}
\newcommand{\eea}{\end{eqnarray}}
\newcommand{\etal}{ {\it{et. al.}}}

\begin{document}

\title{Auxiliary-field quantum Monte Carlo study of first- and 
   second-row post-d elements}

\author{W. A. Al-Saidi, Henry Krakauer, and Shiwei Zhang}

\affiliation
{Department of Physics, College of William and Mary,
Williamsburg, VA 23187-8795}
\date{\today}

\begin{abstract}

A series of calculations for the first- and second-row post-d elements
(Ga-Br and In-I) are presented using the phaseless auxiliary-field quantum Monte
Carlo (AF QMC) method. 
This method is formulated in a Hilbert space defined by any 
chosen one-particle basis, and maps the many-body problem into a linear
combination of independent-particle solutions with
external auxiliary fields. The phase/sign problem is handled approximately by the phaseless formalism
 using a
trial wave function, which in our calculations was chosen to be the Hartree-Fock
solution.  We
used the consistent correlated basis sets of Peterson and coworkers,
which employ a small core relativistic pseudopotential. 
The AF QMC results are compared with experiment
and with those from density-functional (GGA and B3LYP) and 
coupled-cluster CCSD(T) calculations.
The AF QMC total energies agree with CCSD(T) to within a few
milli-hartrees across the systems and over several basis sets.  The
calculated atomic electron affinities, ionization energies, and
spectroscopic properties of dimers are, at large basis sets, in
excellent agreement with experiment.

\end{abstract}
\maketitle

\section{Introduction}

We recently extended the phaseless auxiliary-field quantum Monte Carlo
(AF QMC) approach \cite{zhang_krakauer} to any single-particle basis, and
applied it to the study of molecular systems with Gaussian basis sets
\cite{gafqmc}. The calculated all-electron total energies of many
first-row atoms and molecules at their equilibrium geometries show
typical systematic errors of no more than a few milli-hartrees (mE$_h$)
compared to exact results. This is roughly comparable to that of
CCSD(T), coupled-cluster with single and double excitations plus an
approximate treatment of triple excitations.  For stretched bonds in
H$_2$O \cite{gafqmc} as well as N$_2$ and F$_2$ \cite{gafqmc2}, the phaseless AF QMC
exhibits better overall accuracy and a more uniform behavior than
CCSD(T) in mapping the potential energy curve.

In this paper, we apply the new method to heavier systems and 
present a systematic study of several properties of the first- and
second-row post-d elements (Ga-Br and In-I). Our goal is to
systematically benchmark the new method in different environments and
to compare it to experiment, as well as to a high level correlation method such
as CCSD(T). 

Throughout this work, we used the consistent correlated basis sets of
Peterson and co-workers, which were introduced recently for the
first-, second-, and third-row non-transition metal post-d elements
\cite{peterson1,peterson2}. These are denoted by cc-pVnZ-PP (with $n=$
D, T, Q, 5) \cite{basis_sets_web}, and systematically converge to the
complete basis set limit much like the correlation consistent basis
sets of Dunning and co-workers for light atoms
\cite{dunning1,dunning2}.

The cc-pVnZ-PP basis sets are not all-electron basis sets. They 
employ a small-core relativistic pseudopotential where the $(n-1)spd$
semi-core electrons are treated explicitly in the valence space, and
only the [Ne], [Ar]$3d^{10}$, and [Kr]$4d^{10}\,4f^{14}$ cores are
replaced by pseudopotentials for the first-, second-, and third-row
elements, respectively. This is to be contrasted with large-core
pseudopotentials where the $(n-1)spd$ semi-core electrons are also
removed by the pseudopotential procedure. Large-core pseudopotentials
retain a smaller number of valence electrons, but this is
done at the expense of decreasing the transferability of the
pseudopotentials. For example, it was shown that large-core
pseudopotentials would lead to an overestimation of the correlation
energy of the valence electrons compared to all-electron results by as
much as $10$\%. \cite{dolg_large_ecp}

For heavier atoms, relativistic effects become more important. A
simple and straightforward way to include the scalar relativistic
effects
is through the use of relativistic pseudopotentials, which are
constructed from fully relativistic all-electron atomic calculations.
Thus, in these systems, relativistic pseudopotentials not only
reduce the number of electrons and basis size, but more importantly,
help to include scalar relativistic effects in a non-relativistic
type calculation. The pseudopotentials introduced with cc-pVnZ-PP are
of very high quality, as was verified by comparing relativistic
all-electron calculations with those obtained using these
pseudopotentials \cite{peterson1,peterson2}.

In electronic structure calculations, 
density functional methods \cite{kohn_nobel} are by far the most
widely used.  These methods have low
computational cost, and have allowed accurate predictions of many
properties. In the chemistry community, hybrid exchange correlation functionals are
often used, with B3LYP perhaps the most popular choice.
These approaches are not without
problems, however, especially for systems with stronger electron correlation
effects. 
There is so far no systematic way to improve upon the 
current density functionals to allow sufficiently accurate and robust
calculations over broad
classes of systems.

The exact full configuration interaction (FCI) method scales exponentially with
the number of electrons and is limited to the smallest systems. 
Among the correlated methods, CCSD(T) is the most well-established.
For molecules, CCSD(T) is generally
very accurate near the equilibrium geometry, but its accuracy deteriorates
as bonds are stretched
\cite{H2O_Olsen,garnet_N2,Musial_bartlett_F2}. The application of
CCSD(T) to larger systems is also severely
limited by its computational cost, which grows as $N^7$ in basis size,
as well as by its large memory and disk space requirements.

Quantum Monte Carlo (QMC) methods are attractive due to their
favorable scaling with system size. (In our present implementation
with Gaussian basis sets, the method scales as $N^3$ to $N^4$ with basis
size).  QMC methods
approach the solution of the problem through a
stochastic sampling of the many-body wave function. One price is that
the statistical error only decays as the
square root of the computer time. 
A far more serious problem, if uncontrolled, arises for Fermion wave functions
in the form of divergent statistical fluctuations. This is the well-known 
sign/phase problem \cite{anderson_fixednode,QMC_rmp,zhang_review}.

No formal solution has been found for the sign/phase problem. However,
there are several QMC methods that control it. The most
commonly used QMC method for continuum systems is the real space diffusion
Monte Carlo (DMC) method, which
uses the fixed-node approximation \cite{anderson_fixednode} to control
the sign problem. DMC has been applied to calculate many
properties of solids and molecules \cite{QMC_rmp}. An alternative, the 
auxiliary-field QMC (AF QMC) method, which is relatively new in {\em ab initio} calculations, uses the phaseless formalism to control
the phase problem \cite{zhang_krakauer}. This method has been applied
to $sp$-bonded atoms, molecules, and solids
\cite{zhang_krakauer,zhang_krakauer2,cherry} and transition metal
molecules TiO and MnO \cite{alsaidi_tio_mno}, using a planewave basis
and pseudopotentials. It was also applied recently using a Gaussian
basis to a variety of first-row atoms and molecules \cite{gafqmc}.

Compared with previous efforts \cite{Baer,Silvestrali} on realistic 
electronic systems
using the standard auxiliary-field formalism \cite{BSS,Koonin},
the phaseless AF QMC method overcomes the poor (exponential) scaling 
with system size and projection time, and 
has statistical errors that are 
well-behaved. 
The systematic error from the phaseless approximation has proved 
small in the applications above and, 
as we will show in the present study, 
in the heavier post-d systems. 
All of these calculations have required 
in the phaseless approximation 
only the Hartree-Fock or 
density-functional solution as input.

The phaseless AF QMC method thus provides a many-body framework for
solving the Schr\"odinger equation written in a Hilbert space spanned
by some fixed one-particle basis, and systematically includes
correlation effects by building stochastic ensembles of
independent-particle solutions.  The method reduces the many-body
calculations to manipulations of single-particle orbitals, which are
therefore shared with typical electronic structure methods.  For
example, the localized basis approach used in this paper imports the
one- and two-body Hamiltonian matrix elements directly from standard
quantum chemistry calculations.  This is appealing for quantum
chemistry, especially with the advanced status of basis sets which are
tailored towards correlated methods
\cite{dunning1,dunning2,peterson1,peterson2}.  For planewave basis
 sets, the AF QMC methodology can take full advantage of well-established techniques
 used by independent particle methods, such as fast Fourier
 transforms. Pseudopotentials or effective core
 potentials can be treated straightforwardly with either basis. For the systems
 studied here, the use of standard Gaussian basis sets resulted in a
significant efficiency gain compared to planewave pseudopotential
calculations
\cite{zhang_krakauer,gafqmc,zhang_krakauer2,cherry,alsaidi_tio_mno}.

The rest of the paper is organized as follows. The phaseless AF QMC
method is first briefly reviewed in the next section.  In
Sec.~\ref{sec:results}, we present and discuss the results of our
calculations of the electron affinity and ionization energy of  first-
and second-row post-d elements. In Sec.~\ref{sec:dimers}, we will show
our results for the dissociation energies, equilibrium bondlengths,
and angular frequency of vibrations of three representative post-d
dimers,  As$_2$, Br$_2$, and Sb$_2$.  Finally, in Sec.~\ref{sec:summary}
we conclude with a brief summary.

\section{The phaseless AF QMC Method} \label{sec:method} 

The phaseless auxiliary-field quantum Monte Carlo method belongs to the class of
stochastic projection methods for evolving the imaginary-time
Schr\"odinger equation,
\be
-\frac{ \partial  \left | \Psi (\beta)\right\ra  } { \partial \beta}=
{\hat H} \left |  \Psi (\beta) \right\ra ,  \label{eq:proj} 
\ee
subject to a boundary condition at $\beta=0$.  For time independent
Hamiltonians, the formal solution of Eq.~(\ref{eq:proj}) is,
\be
\left|\Psi (\beta)\right\ra =  e^{-\beta \hat{H}}\, \left|\Psi
(\beta=0) \right\ra ,  \label{eq:proj_sol}
\ee
where $ \left|\Psi (\beta=0)\right\ra= \left| \Psi_T\right\ra $ 
and $\left| \Psi_T\right\ra$ is a trial wave
function determined from, for example, a mean-field type calculation.
Equation~(\ref{eq:proj_sol}) shows the projective nature of 
Eq.~(\ref{eq:proj}). 
If $\left|\Psi_T\right\ra
$  has a non-zero overlap with the exact ground state of the system,
the excited state contributions of
$\hat{H}$ are continuously projected out from $\left|\Psi(\beta)\right\ra$ with an exponential rate determined by their
separation from the ground state.

${\hat H}$ is the many-body Hamiltonian of the system. For electronic systems,
it can be
written in any one-particle basis as, 
\begin{equation}
{\hat H} ={\hat H_1} + {\hat H_2}
= \sum_{i,j}^N {T_{ij} c_i^\dagger c_j}
   + {1 \over 2} 
\sum_{i,j,k,l}^N {V_{ijkl} c_i^\dagger c_j^\dagger c_k c_l},
\label{eq:H}
\end{equation}
where $N$ is the size of the chosen one-particle basis, and
$c_i^\dagger$ and $c_i$ are the corresponding creation and
annihilation operators.  The one-electron $T_{ij}$ and two-electron
$V_{ijkl}$ matrix elements  depend on the chosen basis. 

\begin{table*}[tbh]
\caption{ Total energies for the first-row Ga-Br post-d atoms and negative ions as
  calculated using UHF, B3LYP, CCSD(T), and QMC methods with an
  aug-cc-pVDZ-PP basis set.  QMC/UHF shows the QMC total energy with
  the UHF trial wave function, and QMC/B3LYP shows the corresponding
  value with the B3LYP trial wave function. The variational energy of
  the B3LYP Slater determinant is shown under B3LYP/VAR. Energies are
  in hartrees, and statistical errors  are on the last digit and
  are shown in parentheses.}
\begin{ruledtabular}
\begin{tabular}{ldddddd}
\multicolumn{1}{c}{Atom} & \multicolumn{1}{c}{UHF}  &
\multicolumn{1}{c}{ B3LYP/VAR}  & \multicolumn{1}{c}{B3LYP}& 
\multicolumn{1}{c}{CCSD(T)}  &  \multicolumn{1}{c}{QMC/UHF} &  \multicolumn{1}{c}{QMC/B3LYP}  \\
\hline
 Ga &      -258.291\,230  &  -258.281\,621  &  -259.491\,652     &  -258.361\,690  & -258.358\,9(2) & -258.358\,8(2)\\
 Ga$^-$ &  -258.289\,065 &  -258.277\,828   &  -259.506\,050    &  -258.371\,668  & -258.367\,7(2) & -258.367\,8(2)\\
\\
 Ge     &  -293.341\,749  & -293.333\,305   &  -294.568\,042    &  -293.428\,255   & -293.425\,4(2) & -293.425\,4(2)\\
 Ge$^-$ &  -293.372\,909  & -293.363\,549   & -294.615\,432     &  -293.474\,109   & -293.472\,4(4) & -293.473\,4(3)\\
\\
 As     &  -331.198\,045   & -331.191\,603 &  -332.456\,709    &  -331.299\,321  & -331.299\,9(1) & -331.299\,5(1)\\
 As$^-$ &  -331.185\,601  &  -331.176\,382   &  -332.490\,511    &  -331.312\,824   & -331.310\,2(2) & -331.311\,0(2)\\
\\
 Se     &  -371.846\,027  & -371.838\,683   &  -373.170\,972    &  -371.966\,003   & -371.963\,7(3) & -371.963\,6(2)\\
 Se$^-$ &  -371.881\,023  & -371.872\,353   &  -373.250\,135    &  -372.030\,137 & -372.028\,5(4)  & -372.028\,3(4)\\
\\
 Br     &  -415.474\,798 & -415.467\,803   &  -416.861\,938     &  -415.614\,837  & -415.613\,1(3) & -415.613\,3(2)\\
 Br$^-$ &  -415.564\,018 & -415.556\,914   &  -416.991\,753     &  -415.735\,061  & -415.736\,3(4)  & -415.736\,8(4)\\
\end{tabular}
\label{table_augdz_b3lyp_trialwfn}
\end{ruledtabular}
\end{table*}

Given the general form of the many-body Hamiltonian of 
Eq.~(\ref{eq:H}), the imaginary-time propagator $e^{-\tau {\hat H}}$  of
Eq.~(\ref{eq:proj_sol}) can be written using Trotter decomposition as 
$
e^{-\tau {\hat H}}\doteq e^{-\tau {\hat H_1}/2} e^{-\tau {\hat H_2}}
e^{-\tau {\hat H_1}/2} 
$
for sufficiently small time-step, $\tau$. This would result in a
Trotter time-step error, 
which can
be eliminated by an extrapolation to $\tau =0$ with multiple calculations. The central
idea in the AF QMC method is the use of the Hubbard-Stratonovich (HS)
transformation \cite{HS}:
\begin{equation}
   e^{-\tau{\hat H_2}}
= \prod_\alpha \Bigg({1\over \sqrt{2\pi}}\int_{-\infty}^\infty
d\sigma_\alpha \,
            e^{-\frac{1}{2} \sigma_\alpha^2}
           e^{\sqrt{\tau}\,\sigma_\alpha\,
\sqrt{\zeta_\alpha}\,{\hat v_\alpha}} \Bigg).
\label{eq:HStrans1}
\end{equation}
Equation~(\ref{eq:HStrans1}) introduces 
{\emph{one-body operators}} ${\hat v_\alpha}$ which can be
defined generally for any two-body operator by writing the latter in a
quadratic form, such as ${\hat H_2} = - {1\over 2}\sum_\alpha
\zeta_\alpha {\hat v_\alpha}^2$, with $\zeta_\alpha$ a real
number. The many-body problem as defined by $\hat{H_2}$ is now mapped
into a linear combination of non-interacting problems defined by
${\hat v_\alpha}$, interacting with external auxiliary
fields. Averaging over different auxiliary-field configurations is then performed
by Monte Carlo (MC) techniques.
Formally, this leads to a representation of 
$\left|\Psi (\beta)\right\ra$ as a linear combination of an ensemble of 
Slater determinants, $\{\,\left|\phi (\beta)\right\ra\,\}$.
The orbitals of each $\left|\phi (\beta)\right\ra$ are written in terms of 
the chosen one-particle basis and stochastically evolve with $\beta$. 

However, except for special cases (e.g., the Hubbard model with
on-site interaction), the two-body interactions will require
\cite{zhang_krakauer} complex
one-body operators $\v_op \equiv \{\,{\sqrt\zeta_\alpha \hat v_\alpha}\,\}$.  As a
result, the orbitals in $\left|\phi(\beta)\right\ra$ will
become complex for $\beta > 0$.  For large projection times $\beta$, the
phase of each $\left|\phi(\beta)\right\ra$ becomes random, and
the MC representation of $\left|\Psi (\beta)\right\ra$ becomes
dominated by noise. This leads to the phase problem and the divergence
of the fluctuations. The phase problem is of the same origin as the
sign problem that occurs when the one-body operators $\v_op$ are real,
but is more severe because, instead of a $+\left|\phi(\beta)\right\ra$
and $-\left|\phi(\beta)\right\ra$ symmetry \cite{Zhang}, there
is now an infinite set $\{ e^{i\theta} \left|\phi(\beta)\right\ra, \theta \in [0,2\pi) \}$ among which the Monte Carlo
sampling cannot distinguish.

The phaseless AF QMC method \cite{zhang_krakauer} used in this paper
controls the phase/sign problem in an approximate manner using a trial
wave function. The method recasts the imaginary-time path integral as
a branching random walk in Slater-determinant space
\cite{Zhang,zhang_krakauer}.  It uses 
a {\em complex} importance function, the overlap $\langle
\Psi_T|\phi( \beta)\rangle$, to construct phaseless random walkers,
$|\phi( \beta)\ra/\langle \Psi_T|\phi( \beta)\rangle$, which are invariant under a
phase gauge transformation. The resulting two-dimensional diffusion
process in the complex plane of the overlap $\langle
\Psi_T|\phi( \beta)\rangle$ is then approximated as a diffusion process in one
dimension. We comment that the phaseless constraint confines the random
walk in Slater determinant space according to its overlap with a trial
wave function. This overlap is a global property and is different from
a nodal condition in real electronic configuration space
\cite{zhang_review}.  Thus, the phaseless approximation can behave
differently from the fixed node approximation in DMC.

The ground-state energy computed with the so-called mixed estimate is
approximate and not variational in the phaseless method.  The error
depends on $|\Psi_T\rangle$, vanishing when $|\Psi_T\rangle$ is
exact. This is the only error in the method that cannot be eliminated
systematically.  In tests to date
\cite{zhang_krakauer,gafqmc,zhang_krakauer2,alsaidi_tio_mno,cherry},
the trial wave function has been a single Slater determinant taken
directly from mean-field calculations, and the systematic error has
proved quite small.  For example, in the first-row elements and
molecules, the QMC energies agree to within a few mE$_h$
\cite{gafqmc}, with exact values, using Hartree-Fock solutions as
trial wave functions. 

\begin{table*}[htb]
\caption{ The electron affinity of the first- and second-row post-d
  elements calculated using UHF, GGA, B3LYP, CCSD(T), and the present
  QMC methods. In CCSD(T)$^*$ only the valence $ns$ and $np$ electrons
  are correlated. Results for three basis sets, aug-cc-pVDZ-PP, aug-cc-pVTZ-PP,
  and aug-cc-pVQZ-PP, are shown.  Experimental values are from
  Ref. \cite{exp_EA} with spin-orbit effects approximately removed 
\cite{exp_EA_moore} except for Se and Te, where there is no appropriate 
 experimental data. The mean absolute error (m.a.e.) from
  the experimental data is also shown for each method and basis
  set (the average error on experimental data is 1~kcal/mol). QMC statistical errors are on the last digit and are shown in
  parentheses. Energies are in kcal/mol.  }
\begin{ruledtabular}
\begin{tabular}{lddddddc}
  & \multicolumn{1}{c}{UHF} & \multicolumn{1}{c}{GGA} & \multicolumn{1}{c}{B3LYP} & \multicolumn{1}{c}{CCSD(T)$^*$} & \multicolumn{1}{c}{CCSD(T)} & \multicolumn{1}{c}{QMC} &   \multicolumn{1}{c}{Expt.}  \\ 
\hline
aug-cc-pVDZ-PP\\
 Ga &  -1.36  &  10.845   &  9.03    &  6.13   &  6.26   &  5.9(1)   &   $6\pm3 $  \\
 Ge &  19.55  &  32.18   &  29.74   &  28.49   &  28.77   & 30.1(1)   &   $31.20\pm0.07  $ \\
 As &  -7.81   &  18.36   &  21.21    &  8.32   &  8.47   &  7.2(2)   &  $ 16.0\pm0.7 $  \\
 Se &  21.96   &  48.13   &  49.68   & 40.16   &  40.25   &  40.8(2)   &   $46.60\pm0.01  $ \\
 Br &  55.99   &  80.38   &  81.46   &  75.33   &  75.44   &  77.5(2)   &   $81.11\pm0.07  $ \\
 In &  1.90    &  12.20   &  10.96   &  8.44   &  8.65   &  8.3(2)   &   $7\pm5 $  \\
 Sn &  22.33   &  32.64   &  30.87    &  29.91   &  30.32   &  31.5(1)   &  $ 32.79\pm0.09 $  \\
 Sb &  -2.40   &  20.97   &  23.96   &  11.72   &  11.86   &  10.7(1)   &   $18.7\pm1.2   $\\
 Te &  24.98   &  47.81   &  49.58   &  40.51   &  40.63   &  40.6(2)   &  $ 45.45\pm0.01 $  \\
\smallskip
 I &  55.33    &  76.15   &  77.44   &  71.51   &  71.70   &  73.7(1)   &   $77.791\pm0.002$  \\
m.a.e. & 17.2 & 2.2    &  2.9   & 4.5 & 4.4 & 3.9(1) & \\
\\
aug-cc-pVTZ-PP\\
Ga &  -1.47   &  11.03   &  9.20  &  7.24   &  7.33   &  6.6(2)   &  $ 6 \pm 3  $ \\
 Ge &  19.31   &  32.19   &  29.73  &  30.45   &  30.81   &  31.5(4)   &   $31.20 \pm 0.07  $ \\
 As &  -6.89    &  19.24   &  22.01  &  14.24   &  14.26   &  12.2(5)   &   $16.0 \pm 0.7 $  \\
 Se &  21.38   &  47.93   &  49.56    &  43.96   &  44.12   &  44.2(4)   &  $ 46.60 \pm 0.01 $  \\
 Br &  54.01    &  79.27   &  80.50   &  76.96   &  77.20   &  80.7(6)   &  $ 81.11 \pm 0.07 $  \\
 In &  1.74    &  12.24   &  10.98   &  9.58   &  9.76   &  10.2(2)   &   $7 \pm 5 $  \\
 Sn &  22.09    &  32.57   &  30.77   &  31.94   &  32.47   &  33.2(3)   & $  32.79 \pm 0.09  $ \\
 Sb &  -1.41   &  21.77   &  24.63  &  17.92  &  17.94   &  15.0(7)   &  $ 18.7 \pm 1.2  $ \\
 Te &  24.37     &  47.51   &  49.30  &  44.40   &  44.67   &  45.5(5)   &  $ 45.45 \pm 0.01  $ \\
\smallskip
 I &  53.10    &  74.79   &  76.19   &  72.90   &  73.33   &  77.3(6)   &  $ 77.791 \pm 0.002  $ \\
m.a.e. & 17.6 & 2.6 & 3.2 & 2.1 &  1.9 & 1.5(4) &  \\ 
\\
aug-cc-pVQZ-PP\\
 Ga &  -1.49   &  11.32   &  9.36   &  7.48   &  7.40   &  6.3(6)   &  $6 \pm 3$  \\
 Ge &  19.24   &  32.38   &  29.81   &  30.85   &  31.07   &  31.5(8)   &   $  31.20 \pm 0.07 $  \\
 As &  -6.78    &  19.60   &  22.22  &  15.95   &  16.00   &  13.7(6)   &  $ 16.0 \pm 0.7  $ \\
 Se &  21.34   &  48.09   &  49.62    & 46.03   &  46.17   &  47.1(8)   &  $ 46.60 \pm 0.01 $  \\
 Br &  53.89   &  79.32   &  80.48   &  79.43   &  79.61   &  80.7(8)   &   $ 81.11 \pm 0.07 $  \\
 In &  1.73   &  12.63   &  11.24    &  9.86   &  9.94   &  10.4(7)   &  $ 7 \pm 5  $ \\
 Sn &  22.00    &  32.78   &  30.87   &  32.39   &  32.85   &  32.7(4)   &  $ 32.79 \pm 0.09  $ \\
 Sb &  -1.19   &  22.35   &  25.05   &  19.80   &  20.04   &  18.3(5)   &   $18.7 \pm 1.2 $  \\
 Te &  24.37     &  47.81   &  49.52   & 46.68   &  47.06   &  46.4(6)   &   $45.45 \pm 0.01  $ \\
\smallskip
 I &  52.97   &  74.95   &  76.29   &  75.65  &  76.08   &  79.1(6)   &   $77.791 \pm 0.002  $ \\
m.a.e. & 17.6&  2.8 &   3.3 &  1.2  & 1.1 &   1.0(6) &   
\end{tabular}
\label{table_dz_tz_qz_EA}
\end{ruledtabular}
\end{table*}

\section{Results and discussion: Atomic properties}\label{sec:results}

As mentioned, our AF QMC method uses a trial wave function to control
the sign/phase problem. In general, the trial wave function has to be
in the form of a Slater determinant or a linear combination of Slater
determinants. In our previous study, we found that using the
unrestricted Hartree-Fock (UHF) solution  leads to QMC energies which
are in better agreement with exact energies than those obtained using
the restricted Hartree-Fock (RHF) Slater determinant. This was the
case even with singlets.  QMC results obtained using density
functional generated trial wave functions lead to the same energies as
Hartree-Fock generated wave functions, within statistical errors
\cite{gafqmc}.

We find the same insensitivity of the AF QMC results 
to the choice of trial wave function in post-d systems.
In
Table~\ref{table_augdz_b3lyp_trialwfn}, we summarize our comparisons
using the aug-cc-pVDZ-PP basis set for the first-row Ga-Br post-d
elements. We carried out the QMC calculations using both UHF and unrestricted B3LYP
trial wave functions. As a measure of the difference between the two
Slater determinants, we show the variational energy, $\la \Psi_T |{\hat H}|
\Psi_T \ra$, of the B3LYP trial Slater determinant $|\Psi_T\ra$, which
is higher, of course, than the UHF energy.  Both QMC values are the same
within statistical errors. 

For the rest of our study, we will always use the UHF determinant as the
trial wavefunction when it exists, and the RHF solution otherwise. 
All of our QMC calculations are performed
with several Trotter time-steps, and we report only  the
extrapolated values.

Table~\ref{table_augdz_b3lyp_trialwfn} shows that the QMC 
and CCSD(T) energies agree to within a few
mE$_h$. 
This
agreement is similar to that found previously for the lighter systems
comprised of first-row atoms and molecules \cite{gafqmc}. 
In that study, CCSD(T) and QMC energies were found to be roughly comparable in their
agreement with exact values \cite{gafqmc}. 
(Both CCSD(T) and
the present QMC method are non-variational.)

All of the Hartree-Fock, density functional, and coupled cluster
calculations were carried out using NWCHEM \cite{nwchem}. Some of them
were 
also verified using Gaussian~98 \cite{gaussian98}. For open shell
systems, we used the UHF as the reference state for the coupled cluster
calculations. For closed shell systems, we used the RHF reference
states, unless otherwise specified. 
[For example, the dimers studied in Sec.~\ref{sec:dimers} are singlets,
but their potential energy curves were studied with both RCCSD(T) and UCCSD(T).]
All of our calculations were performed using non-relativistic
methods. 

Finally, all of the QMC calculations in this paper correlated  all the electrons 
in the wave function, {\em i.e.}, no frozen-core approximation was made.
These results are compared with
those obtained using CCSD(T) with all the valence electrons correlated,
as well as with CCSD(T) results obtained with only the valence $ns$
and $np$ electrons correlated (denoted by CCSD(T)$^*$).
We note that the consistent correlation
cc-pVnZ-PP basis sets used here are in fact optimized for the correlation energy of the
valence electrons only \cite{peterson1,peterson2}, but this does not affect the benchmarking of our
results against CCSD(T).

\begin{table}[t]
\caption{Energies for the first- and second-row post-d group elements
  as calculated using UHF, CCSD(T), and the present QMC
  methods. Calculations are done with aug-cc-pVQZ-PP basis sets. The
  average absolute difference between QMC and  CCSD(T) values is $(1.2
  \pm 7)$ mE$_h$. QMC statistical errors  are on the last digit and
  are shown between parentheses. Energies are in hartrees.}
\begin{ruledtabular}
\begin{tabular}{ldddd}
\multicolumn{1}{c}{}  & \multicolumn{1}{c}{UHF}  &
\multicolumn{1}{c}{CCSD(T)}  &  \multicolumn{1}{c}{QMC}  \\
\hline
Ga &  -258.303\,227       &  -258.571\,831   &  -258.572\,2(7) \\
 Ga$^-$ &  -258.300\,850     &  -258.583\,627   &  -258.582\,2(5) \\
 \\
 Ge &  -293.353\,889      &  -293.632\,818   &  -293.633\,0(9) \\
 Ge$^-$ &  -293.384\,544      &  -293.682\,326   &  -293.683\,1(9) \\
 \\
 As &  -331.211\,170      &  -331.508\,405   &  -331.510\,9(6) \\
 As$^-$ &  -331.200\,364       &  -331.533\,906   &  -331.532\,6(8) \\
 \\
 Se &  -371.864\,097       &  -372.193\,795   &  -372.192\,2(9) \\
 Se$^-$ &  -371.898\,102    &  -372.267\,364   &  -372.267\,3(8) \\
 \\
 Br &  -415.493\,211      &  -415.859\,089   &  -415.858\,6(6) \\
 Br$^-$ &  -415.579\,084     &  -415.985\,953   &  -415.987\,0(9) \\
 \\
 In &  -189.210\,636       &  -189.436\,098   &  -189.434\,3(6) \\
 In$^-$ &  -189.213\,399       &  -189.451\,931   &  -189.450\,9(9) \\
 \\
 Sn &  -213.340\,624      &  -213.580\,057   &  -213.580\,3(5) \\
 Sn$^-$ &  -213.375\,689     &  -213.632\,412   &  -213.632\,5(3) \\
 \\
 Sb &  -239.277\,991       &  -239.535\,458   &  -239.536\,3(6) \\
 Sb$^-$ &  -239.276\,089       &  -239.567\,392   &  -239.565\,4(6) \\
 \\
 Te &  -267.006\,904     &  -267.286\,525   &  -267.285\,2(8) \\
 Te$^-$ &  -267.045\,732   &  -267.361\,526   &  -267.359\,2(5) \\
 \\
 I &  -296.659\,645     &  -296.990\,114   &  -296.987\,3(4) \\
 I$^-$ &  -296.744\,059      &  -297.111\,355   &  -297.113\,4(8) \\
\end{tabular}
\label{table_aug_qz_EA}
\end{ruledtabular}
\end{table}

\begin{table*}[t]
\caption{The first ionization potential of first- and second-row
  post-d elements. In CCSD(T)$^*$ only the valence $ns$ and $np$
  electrons are correlated. Three basis sets cc-pVDZ-PP, cc-pVTZ-PP,
  and cc-pVQZ-PP are used.  Experimental values are from
  Ref. \cite{exp_EA} with spin-orbit effects approximately removed
  \cite{exp_EA_moore}. The mean absolute error (m.a.e.) from the
  experimental data is also shown for each method and basis set.
QMC
  statistical errors  are on the last digit and are shown between
  parentheses. All energies are in eV.
}
\begin{ruledtabular}
\begin{tabular}{lddddddd}
   & \multicolumn{1}{c}{UHF} & \multicolumn{1}{c}{GGA} &
  \multicolumn{1}{c}{B3LYP} & \multicolumn{1}{c}{CCSD(T)$^*$} &  \multicolumn{1}{c}{CCSD(T)} & \multicolumn{1}{c}{QMC} &   \multicolumn{1}{c}{Expt.}  \\ 
\hline
cc-pVDZ-PP\\
 Ga &  5.51   &  5.91   &  5.94 &  5.74     &  5.77   &  5.69(1)   &  5.93  \\
 Ge &  7.41    &  7.83   &  7.82 &  7.69      &  7.71   &  7.72(1)   &  7.95  \\
 As &  9.43   &  9.82   &  9.78&  9.72    &  9.75   &  9.84(1)   &  10.01  \\
 Se &  8.50   &  9.54   &  9.73  &  8.94    &  8.97   &  8.90(1)   &  9.63  \\
 Br &  10.75   &  11.74   &  11.87 &  11.26       &  11.28   &  11.30(1)   &  11.84  \\
 In &  5.20   &  5.57  &  5.63 &  5.39    &  5.43   &  5.37(1)   &  5.60  \\
 Sn &  6.86   &  7.25   &  7.26 &  7.10    &  7.13   &  7.13(1)   &  7.39  \\
 Sb &  8.60   &  8.95   &  8.93   &  8.85    &  8.88   &  8.93(1)   &  9.12  \\
 Te &  7.70   &  8.67   &  8.85  &  8.09   &  8.12   &  8.06(1)   &  8.74  \\
\smallskip
 I &  9.60   &  10.51   &  10.63 &  10.05   &  10.07   &  10.08(1)   &  10.51  \\
m.a.e. & 0.72&  0.09   & 0.11 & 0.39    &   0.36  &  0.37(1)  & \\
\\
cc-pVTZ-PP\\
Ga &  5.53   &  5.91   &  5.96  &  5.87     &  5.89   &  5.79(3)   &  5.93  \\
 Ge &  7.41   &  7.84   &  7.84  &  7.86    &  7.86   &  7.88(3)   &  7.95  \\
 As &  9.42    &  9.84   &  9.80 & 9.94    &  9.94   &  10.06(2)   &  10.01  \\
 Se &  8.55   &  9.60   &  9.79  & 9.39     &  9.40   &  9.30(3)   &  9.63  \\
 Br &  10.74  &  11.76   &  11.90 & 11.62     &  11.62   &  11.63(1)   &  11.84  \\
In &  5.21   &  5.58  &  5.64    &  5.47    &  5.54   &  5.47(3)   &  5.60  \\
 Sn &  6.86    &  7.26   &  7.27  &  7.21   &  7.27   &  7.29(2)   &  7.39  \\
 Sb &  8.60   &  8.97   &  8.95  &  9.01    &  9.07   &  9.13(2)   &  9.12  \\
 Te &  7.75   &  8.73   &  8.90  &  8.49    &  8.54   &  8.46(3)   &  8.74  \\
\smallskip
 I &  9.58   &  10.53   &  10.66 &  10.34    &  10.39   &  10.38(2)   &  10.51  \\
m.a.e. & 0.71 & 0.07 &   0.12 & 0.15     & 0.12    & 0.14(2) & \\ 
\\
cc-pVQZ-PP\\
Ga &  5.53   &  5.93   &  5.96  &  5.84     &  5.91   &  5.77(3)   &  5.93  \\
 Ge &  7.40   &  7.85   &  7.84  &  7.83    &  7.88   &  7.85(3)   &  7.95  \\
 As &  9.41    &  9.84   &  9.80 &  9.92    &  9.97   &  10.04(2)   &  10.01  \\
 Se &  8.54   &  9.60   &  9.78  &  9.48   &  9.53   &  9.42(3)   &  9.63  \\
 Br &  10.71   &  11.75   &  11.88 &  11.71    &  11.75   &  11.73(5)   &  11.84  \\
In &  5.22    &  5.59   &  5.65   &  5.49 &  5.58   &  5.52(4)   &  5.60  \\
 Sn &  6.85   &  7.26   &  7.27   & 7.23   &  7.31   &  7.32(2)   &  7.39  \\
 Sb &  8.58    &  8.96   &  8.94  &  9.03  &  9.11   &  9.17(2)   &  9.12  \\
 Te &  7.73    &  8.73   &  8.90   & 8.61    &  8.69   &  8.60(2)   &  8.74  \\
\smallskip
 I &  9.55   &  10.51   &  10.64  &  10.46  &  10.53   &  10.53(3)   &  10.51 \\
m.a.e. & 0.72 & 0.07  &  0.12  &  0.11   & 0.05    &  0.10(3)& \\
\end{tabular}
\label{table_post_d_ion}
\end{ruledtabular}
\end{table*}

\subsection{Electron affinity}

Atomic electron affinities calculated using different methods are
summarized in Table~\ref{table_dz_tz_qz_EA}, for the aug-cc-pVDZ-PP,
aug-cc-pVTZ-PP, and aug-cc-pVQZ-PP basis sets.  We report the electron
affinities from UHF, GGA \cite{gga}, and hybrid B3LYP \cite{b3lyp}, as
well as those from correlated methods, CCSD(T) and the present QMC. We
also show for comparison the frozen-core CCSD(T)$^*$ results
\cite{peterson1,peterson2}. The experimental electron affinities are
from Ref.~\cite{exp_EA}.  Spin-orbit effects have been approximately removed
by averaging over the atomic multiplets \cite{exp_EA_moore}, except in
Se
and Te for the lack of appropriate experimental data. Both the
independent-electron and correlated results show a smooth convergence
with basis size, and as expected the independent-electron results have
faster convergence.

Results obtained using density functional GGA and hybrid B3LYP methods
are generally in reasonable agreement with experiment.
The values are already converged at the triple-zeta level basis to less
than $1$~kcal/mol. GGA and B3LYP are comparable in
terms of their agreement with the experimental data, with GGA being
slightly better. The worst cases for B3LYP are Sb and As where
the experimental values are overestimated by $\approx 6$~kcal/mol. On the
other hand, GGA overestimates the experimental values by $\approx
3$~kcal/mol. 

The electron affinities obtained using CCSD(T) and QMC methods are in
better agreement with experiment than GGA or B3LYP results at the
aug-cc-pVTZ-PP basis set level, and the agreement reaches chemical
accuracy with the quadruple-$\zeta$ basis set.  CCSD(T) and QMC are comparable, with
deviations of $\approx 1-4$~kcal/mol for all the basis sets studied
and for all systems. Similar agreement is also seen in the total
energies, which are shown in Table~\ref{table_aug_qz_EA}.  The average
absolute difference between QMC and CCSD(T) energies for this set is
$1.2(7)$~mE$_h$, with $2.8(4)$~mE$_h$ the largest difference in the
iodine atom. For the other two basis sets (not shown), the agreement
with CCSD(T) is similar.

Our CCSD(T) calculations for the electron affinities are in good
agreement with the CCSD(T)$^*$ calculations. 
The effects of the frozen-core approximation on the
electron affinity are minimal due to the cancellation of the
frozen-core error between the atom and the ion. 
In addition to
the three basis sets which we used in our study, Peterson and
co-workers report on frozen-core CCSD(T) calculations using
aug-cc-pV5Z-PP basis sets \cite{peterson1,peterson2}. Their electron
affinities using the quadruple and quintuple-$\zeta$ basis sets agree
to less than $1$~kcal/mol with each other.

\begin{table*}[tb]
\caption{
  Summary of the dissociation energies (BE), equilibrium bond
  lengths (R$_e$), and angular frequency of vibrations ($\omega_e$) for
  As$_2$, Br$_2$, and Sb$_2$ post-d dimers using RHF, GGA, B3LYP,
  CCSD(T), and the present QMC methods. In CCSD(T)$^*$, only the
  valence $ns$ and $np$ electrons are correlated
  \cite{peterson1,peterson2}. Two correlation consistent basis sets
  cc-pVDZ-PP and cc-pVTZ-PP are used. 
  Dissociation energies are in kcal/mol, bondlengths
  are in Angstroms, and angular frequencies are in cm$^{-1}$.  The
  experimental values are from Ref.~\cite{ion_ref,sontag} with spin-orbit effects
  approximately removed \cite{exp_EA_moore}. QMC statistical errors 
  are on the last digit and are shown between parentheses.
}
\begin{ruledtabular}
\begin{tabular}{lcccccccc}
  Dimer &           &  RHF & GGA    & B3LYP  &CCSD(T)$^*$ & CCSD(T)& QMC  & Expt. \\
\hline
cc-pVDZ-PP\\
  As$_2$&BE         &8.34 &94.52    & 89.15 & 63.25  & 71.56   & 73.2(4)    & 91.9      \\
        &$R_e$      &2.058 &2.128   & 2.113  &2.145   &  2.135  &  2.136(4)  & 2.103 \\
        &$\omega_e$ &514   & 429    & 444     &414     & 420     & 428(13)   &  429.55 \\
\\
  Br$_2$&BE         &15.37 &54.42    & 45.94      & 36.74       & 39.16  &   42.1(4)  & 52.93   \\
       &$R_e$       &2.290 & 2.331   & 2.335    &  2.341      & 2.334  &  2.321(3)  &  2.281 \\
        &$\omega_e$ &347   & 303     & 305      & 300         & 305    &  314(9)   & 325.31  \\
\\
 Sb$_2$&BE          &$-$9.28 & 72.24   & 67.09     & 40.99    & 49.40 &51.1(4)   & 69.45 \\
        &$R_e$       & 2.443 & 2.516  & 2.509    & 2.554    &2.545  &2.563(8)  &  2.476\\
        &$\omega_e$ &  331  & 276    & 284      & 267      & 259   &225(11)   &  269.62 \\
\\
\\
cc-pVTZ-PP\\
 As$_2$&BE        &12.26 & 96.40   & 91.63   &76.23     &85.76  & 85(1)   &  91.9 \\
       &$R_e$     &2.052 & 2.118   & 2.103   & 2.126    & 2.104 & 2.100(4)&  2.103  \\
       &$\omega_e$&509   & 429     & 444     & 424      & 433   & 427(7) &  429.55 \\
\\
 Br$_2$&BE         &20.66  &57.05  & 48.39     & 44.83   & 48.34   &  52(1)& 52.93\\
        &$R_e$     & 2.272 &2.311  & 2.316     & 2.307    &   2.295 & 2.279(5)& 2.281   \\
        &$\omega_e$& 352   & 312   & 313       & 319     &   322   & 336(12) &   325.31  \\
\\
 Sb$_2$&BE        &$-$5.44 &73.93   &69.05       &52.55    &62.32   & 62(2)    & 69.45\\
       &$R_e$     & 2.436 & 2.505  &  2.491     & 2.532   & 2.506  & 2.512(5)& 2.476 \\
       &$\omega_e$& 327   & 276    &  285       & 267     & 273    & 255(9) & 269.62\\

\end{tabular}
\end{ruledtabular}
\label{table_post_d_be_re_omega}
\end{table*}

\subsection{Ionization energy} \label{sec:ionization}

Ionization energies for the first- and second-row post-d elements using
different methods are shown in Table~\ref{table_post_d_ion}. The
organization is similar to that in Table~\ref{table_dz_tz_qz_EA} for the electron affinities.
The experimental energies are from
Ref. \cite{ion_ref} with spin-orbit effects approximately removed by 
averaging over the multiplets \cite{exp_EA_moore}.

Both the GGA and the hybrid B3LYP density functional methods are in
good agreement with each other and also with the experimental
values. Here also as with the electron affinity, GGA seems to do
better than B3LYP. The results suggest that the independent-electron
ionization potentials are already converged at the double-zeta level
of basis sets.

The correlated ionization energy calculations, namely CCSD(T) and QMC,
are in very good agreement with each other for all the basis sets and
all the atoms. The largest difference between them is for Ga and is
$\approx 0.1$~eV.  The agreement between the total energies of the
positively charged  ions (not shown) obtained
using CCSD(T) and QMC is similar to those shown in
Table~\ref{table_aug_qz_EA}, and the average absolute difference is $<
2$ m$E_h$ over all basis sets. 
The agreement between QMC or CCSD(T) with experiment improves with basis
size as expected, and is very good at the QZ level.

As seen in Table~\ref{table_post_d_ion}, both CCSD(T) calculations with
and without the frozen-core approximation are in excellent agreement
with each other for the ionization energy. Moreover, similar to the
case of the electron affinity, both CCSD(T) results show a smooth
convergence with basis size towards the experimental values.

To our knowledge, our study of the ionization energies is the first of
the post-d elements with the cc-pVnZ-PP basis sets. Our coupled
cluster results are in good agreement with coupled cluster
calculations obtained using all-electron consistent correlation basis
sets \cite{wilson}.  There has also been a study using G2 theory of
the first-row elements, in which the computed ionization energies were
in good agreement with experiment \cite{curtiss_G2}.

\begin{figure}[t]
\includegraphics[width=8cm]{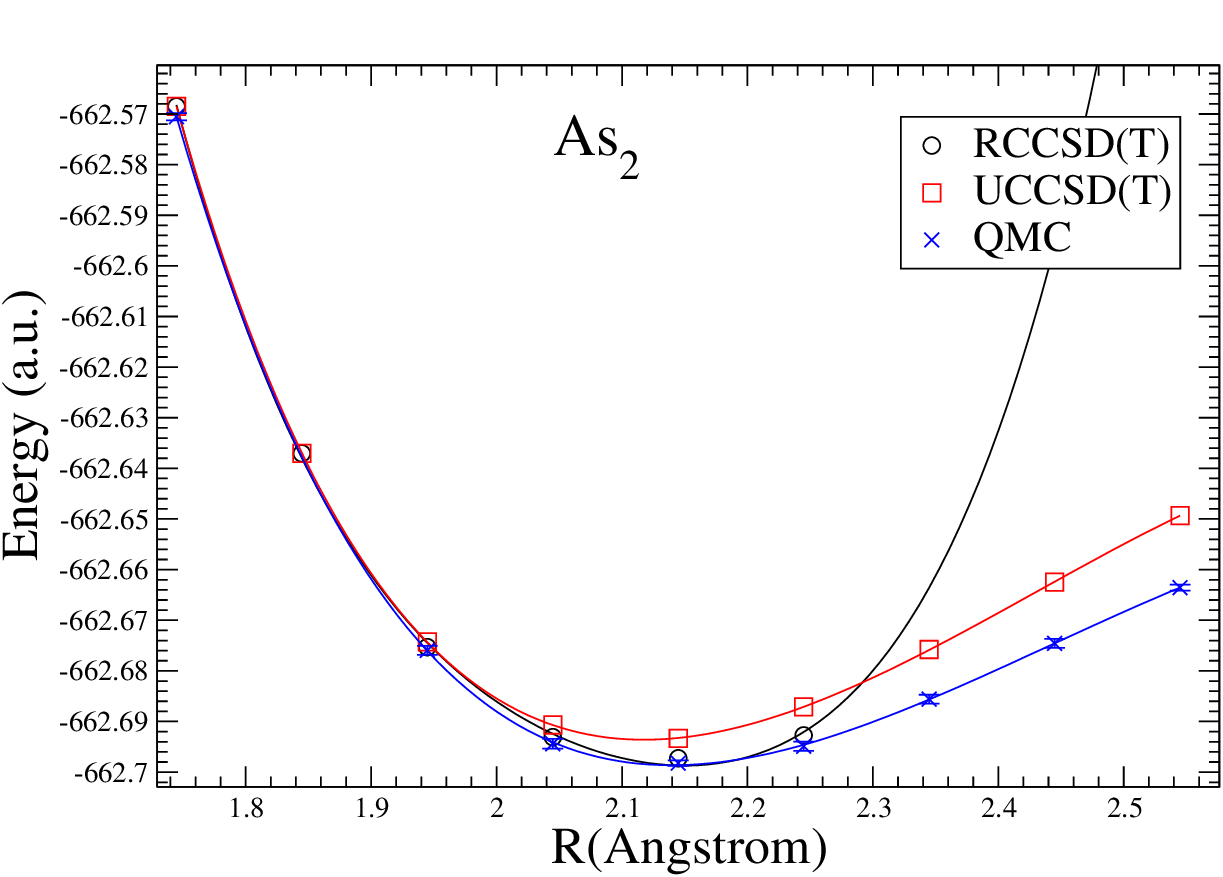} 
\includegraphics[width=8cm]{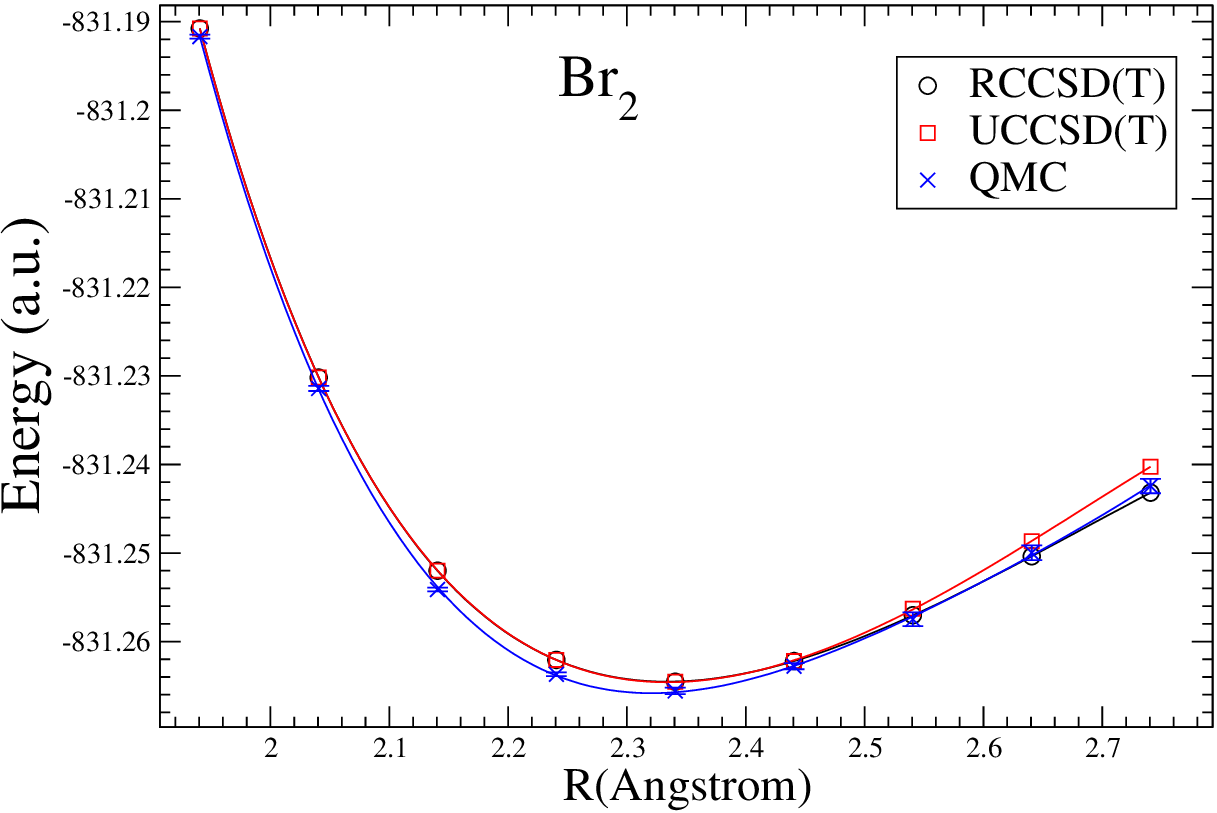} 
\includegraphics[width=8cm]{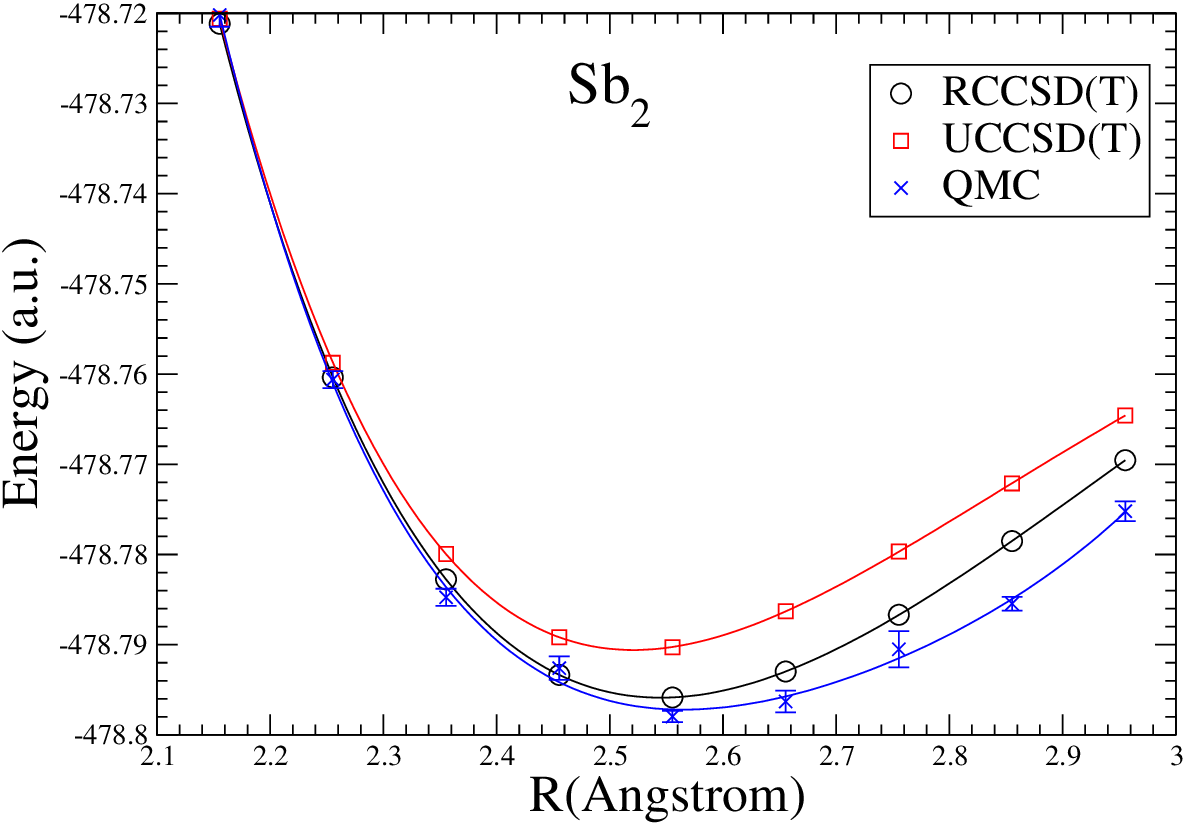}
\caption{Potential energy curves  of As$_2$, Br$_2$, and Sb$_2$
dimers (top to bottom panels, respectively) as obtained using RCCSD(T), UCCSD(T), and the
present QMC methods. Calculations are done using cc-pVDZ-PP basis set. The
solid lines are based on a polynomial fit to the results.}
\label{fig_dimer_dz}
\end{figure}

\begin{figure}[t]
\includegraphics[width=8.2cm]{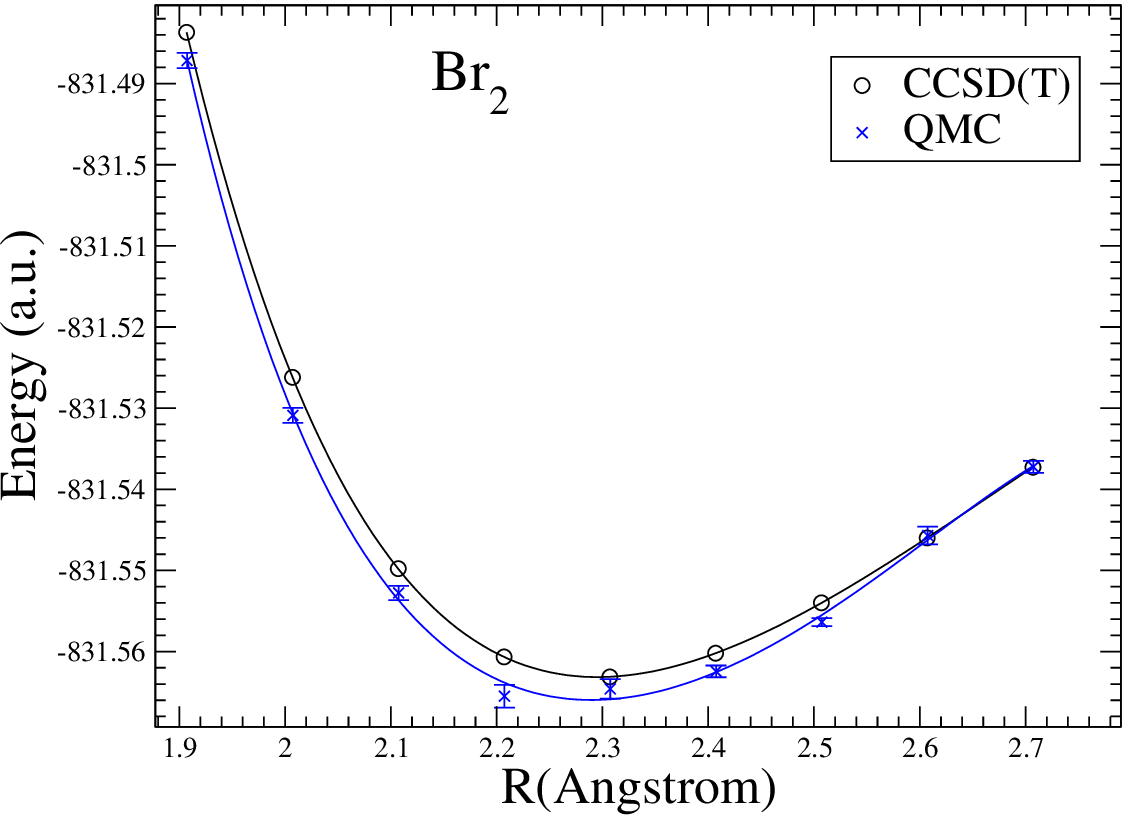} 
\caption{Potential energy surface of Br$_2$ within the cc-pVTZ-PP
basis set, as calculated in RCCSD(T) and QMC. The solid lines are
based on a polynomial fit to the results. }
\label{fig_dimer_tz}
\end{figure}

\section{Results and discussion: dimers}\label{sec:dimers}

We chose As$_2$, Br$_2$, and Sb$_2$ to study,
as representatives of the first- and second-row post-d dimers. In
Table~\ref{table_post_d_be_re_omega}, we summarize our study of their
dissociation energy, equilibrium bondlength, and angular frequency of
vibration using different methods, with the cc-pVDZ-PP and cc-pVTZ-PP
basis sets. Spin-orbit effects in the experimental values
\cite{ion_ref,sontag} are approximately removed \cite{exp_EA_moore}.

Within each method, the equilibrium bondlength and  angular
frequency of vibrations are calculated from a fit to the potential
energy curves of the dimers.  The total energies for 8-9 geometries
over the range $-0.4 \le R-R_e \le 0.7$~Angstroms are fitted by 4-7th
order polynomials, which gave consistent results.
The dissociation energy of the dimers at the RHF, GGA, and B3LYP level
of theories are calculated using the equilibrium bondlength at the
same level of theory and using the same basis. For the correlated
calculations, Table~\ref{table_post_d_be_re_omega} tabulates the calculated
equilibrium bondlength, but the dissociation energies that are shown were
calculated using
the equilibrium geometry optimized with CCSD(T)$^*$, {\em i.e.}, at the
CCSD(T) level of theory with only the $ns$ and $np$ electrons
correlated. This was done to facilitate the comparisons between the
different methods. The dissociation energies at the respective optimized geometries
will change by $1$~kcal/mol or less.

First, we discuss the density functional results. We have verified
that the density functional GGA and hybrid B3LYP dissociation energies
are already converged at the cc-pVTZ-PP basis set. Our dissociation
energy results using cc-pVQZ-PP basis sets (not shown) differ from those reported
at cc-pVTZ-PP by less than $1$~kcal/mol, the 
equilibrium bondlengths 
by less than $0.005$~Angstrom, 
and the angular frequencies 
by less than $1$~cm$^{-1}$.  The density functional dissociation
energies are in reasonable agreement with the experimental values. GGA
overestimates the binding energies by $\approx 4-5$~kcal/mol, while
B3LYP is in good agreement with experiment. The bondlengths are
overestimated with GGA and slightly underestimated with B3LYP. Angular
frequencies in GGA and B3LYP are in better agreement with each other,
with the GGA frequencies being in good agreement with experimental
values.  

We next  comment on the coupled cluster results with and
without the frozen-core approximation.  Contrary to the atomic
properties, the dissociation energy, equilibrium bondlengths, and the
angular frequency are more sensitive to the frozen-core approximation. The
basis sets used in this study are optimized to recover the correlation
energy of the valence electrons, and thus are less effective in accessing the
core-valence correlation effects. In Refs.~\cite{peterson1,peterson2},
the frozen-core CCSD(T) calculations of the spectroscopic properties
of several dimers are in good agreement with the experimental values
with the cc-PV5Z-PP basis sets.

We now focus on the QMC results, and their comparison with CCSD(T) and
experiment. As mentioned, all of the QMC calculations were obtained
using the UHF solution as trial wave function.
QMC and CCSD(T) dissociation energies are in good agreement
with each other. The QMC total energies for the atoms are within
$1$~mE$_h$ of the CCSD(T) values, as reported before in
the study of the ionization energies in Sec.~\ref{sec:ionization}. The
QMC energies of the dimers, on the other hand, are below the CCSD(T)
values by $\approx 1-3$ mE$_h$ for the two basis sets and for all the
studied dimers. This does not necessarily mean that QMC values are in
better agreement with exact values, because these QMC calculations are non-variational (as
is CCSD(T)). In our previous study of the first-row atoms and
molecules where some exact values were available, our comparisons
showed that the exact values often fell between the CCSD(T) and QMC
values.\cite{gafqmc}

The bondlengths and the angular frequency of vibrations obtained with
QMC and CCSD(T) are in less good agreement with each other, compared
to the dissociation energies.  In Fig.~\ref{fig_dimer_dz}, we show the
potential energy curves of As$_2$, Br$_2$, and Sb$_2$ dimers within
the cc-pVDZ-PP basis set, as calculated with QMC, and both RCCSD(T)
and UCCSD(T) which are based on RHF and UHF reference states,
respectively. In As$_2$/cc-pVDZ-PP, we could not obtain the RCCSD(T)
energies for some of the geometries ($R \approx 2.25-2.45$~Angstroms)
due to the lack of convergence, as can be seen in
Fig.~\ref{fig_dimer_dz}.

Figure~\ref{fig_dimer_tz} shows the potential energy surface of Br$_2$
within the cc-pVTZ-PP basis sets, as calculated using CCSD(T) and the
present QMC.  In this case, we show only the RCCSD(T) values, because
we could not obtain a UHF solution in this system for the cc-pVTZ-PP
basis set at the bondlengths shown.

For singlets, the RCCSD(T) is generally more accurate near the
equilibrium bondlength, while UCCSD(T) performs better for larger
bondlengths near the dissociation limit. Both the QMC and RCCSD(T)
energies are in excellent agreement near the
equilibrium bondlength (with QMC slightly lower). However, as the bondlength is stretched,
RCCSD(T) becomes less accurate, and for relatively small bondlength
stretching UCCSD(T) is also inaccurate
\cite{H2O_Olsen,garnet_N2}. QMC, on the other hand, has been shown to give a more
uniform description across the whole potential energy surface in first-row
molecules \cite{gafqmc,gafqmc2}.
Given these results, especially our benchmark study on N$_2$ \cite{gafqmc2}, it seems 
reasonable to speculate that a significant portion of the discrepancy 
at larger bondlengths between QMC and CCSD(T) is due to errors in the latter.
Further study is needed to better establish this.

\section{Summary}
\label{sec:summary}

To further benchmark the recently introduced phaseless auxiliary-field QMC method,
we have applied it 
to heavier systems
using a Gaussian basis. We performed a systematic study of the first-
and second-row non-transition metal post-d elements using the
consistent correlation basis sets cc-pVnZ-PP \cite{peterson1,peterson2}
which employ a small-core relativistic pseudopotentials.
Our results for
the electron affinities and the first ionization potentials of these
elements are in excellent agreement with similar results calculated
using CCSD(T) over double-, triple-, and quadruple-zeta basis sets. The
corresponding energies of the atoms and the ions agree with the
CCSD(T) values, with an average difference of less than $1$~mE$_h$. Our
results obtained using the quadruple zeta basis set are in excellent agreement
with experimental results. 

We also studied the dimers As$_2$, Br$_2$, and Sb$_2$ using cc-pVDZ-PP
and cc-pVTZ-PP basis sets. 
At the triple-zeta level, the calculated spectroscopic properties are
in good agreement with experiment.
Our results for the dissociation energies
are in excellent agreement with 
CCSD(T) within each basis set.  The equilibrium bondlength and the angular
frequency of vibrations are also in good agreement with similar
CCSD(T) results considering the somewhat large statistical errors 
on the QMC results.  The potential energy curves  for all of these
molecules are in good agreement with CCSD(T) for bondlengths smaller
than the equilibrium bondlength. For larger bondlengths, and
especially in As$_2$ and Sb$_2$, the QMC results deviate significantly
($\approx 10$~mE$_h$) from both RCCSD(T) and UCCSD(T). Both
coupled cluster methods are less accurate in these regions
\cite{H2O_Olsen,garnet_N2,Musial_bartlett_F2}, and QMC has been shown
to give better accuracy and a more uniform behavior in H$_2$O, N$_2$, and F$_2$
\cite{gafqmc,gafqmc2}.
The deviations between QMC and CCSD(T) here seem consistent with 
these benchmark results in lighter systems.

\section{Acknowledgments:}

This work is supported by ONR (N000149710049  and
N000140510055), NSF (DMR-0535529), and ARO (48752PH) grants, and by
the DOE computational materials science network (CMSN).  Computations
were carried out at the Center for Piezoelectrics by Design, the
SciClone Cluster at the College of William and Mary, NCSA at UIUC, and
SDSC at UCSD.

\end{document}